\documentclass[aps,pra,twocolumn,showpacs]{revtex4-1}

\usepackage{graphics}

\usepackage{amsmath}
\usepackage{mathtools}
\usepackage{amssymb}
\usepackage{braket}
\usepackage{color}
\usepackage{float}
\usepackage{epsfig}
\usepackage{graphicx,epstopdf}
\usepackage{times}

\usepackage[colorlinks=true, citecolor=blue, urlcolor=blue ]{hyperref}
\input{epsf}
\begin{document}

\definecolor{purple}{RGB}{160,32,240}
\definecolor{orange}{RGB}{255,127,0}

\title{Generalized Geometric Measure of Entanglement for Multiparty Mixed States}
\author{Tamoghna Das, Sudipto Singha Roy, Shrobona Bagchi, Avijit Misra, Aditi Sen(De) and Ujjwal Sen}

\affiliation{Harish-Chandra Research Institute, Chhatnag Road, Jhunsi, Allahabad 211 019, India}

\begin{abstract}

Computing entanglement of an arbitrary bipartite or multipartite mixed state is in general not an easy task as it usually involves complex optimization. 
Here we show that exploiting symmetries of certain  multiqudit mixed states, we can compute a genuine multiparty entanglement measure, the generalized geometric measure, for these classes of mixed states. The chosen states have different ranks and consist of an arbitrary number of parties.

\end{abstract}
\maketitle

\section{Introduction} 
Characterization and quantification of quantum entanglement \cite{HHHHRMP} lies at the heart of  quantum information theory, since its early recognition as ``spooky action at a distance'' \cite{M_Born} in the Einstein-Podolsky-Rosen article \cite{EPR}. 
Moreover, it has been successfully identified as a key resource in several
quantum communication protocols including  superdense coding  \cite{BennetDC}, teleportation \cite{TeleRef}, and quantum cryptography \cite{Crypto}.  
Entanglement has been shown to be a necessary ingredient in studying quantum state tomography \cite{state_tomo}, quantum metrology \cite{Q_metrology}, cooperative
quantum phenomena in many body systems like quantum phase transitions \cite{Phase_transit}, etc. 
Quantification of entanglement is also essential for characterization of successful preparations of quantum states, both in two party and multiparty domains, in the laboratories \cite{Lab_exp}.  

The notion of entanglement is rather well-understood in the bipartite regime, especially for pure states \cite{Ent_entropy,concEof,distillable, LN, Ent_quanti}. While several entanglement measure can be computed for bipartite pure states, the situation for mixed states is difficult, and there are only few 
entanglement measures which can be computed efficiently.  
The logarithmic negativity \cite{LN} can be obtained for arbitrary bipartite states, while the entanglement of formation \cite{concEof,distillable} can be computed for all two-qubit states. The situation becomes complicated even for the pure states 
 when the number of parties  increase. 
However, there have been significant advances in recent times to quantify multipartite entanglement of pure quantum states in arbitrary dimensions \cite{HHHHRMP}. They are broadly classified in two catagories $-$ distance-based measures \cite{GM,Wei_Goldbart,brody,GGM} and  monogamy-based ones \cite{Crypto,Ent_entropy,CKW,discmono}.
On the other hand,   quantifying entanglement for arbitrary multiparty mixed states  is still an arduous task. 
Recently, experiments by using photon polarization \cite{Pan10photon} and ions \cite{ION} have been reported in which multiparty states 
 of the order of ten parties have been created successfully. Such physical implementations demand a general tool to compute multiparty entanglement measures for arbitrary mixed states. Recently there have been notable  advancements in this direction \cite{mixed_ent}. Moreover, when an entanglement measure can only be evaluated for pure states, the entanglement-assisted  study of cooperative phenomena becomes restricted to only a system which is at zero temperature. 

We address here the question of computing the generalized geometric measure (GGM) \cite{GGM}, a genuine multiparty entanglement quantifier, for mixed states. The GGM of pure states has already been computed efficiently in several systems for arbitrary number of parties \cite{GGM2}. 
In this paper, we define the GGM for mixed states via the convex roof.
 To deal with the obstacle of evaluating the convex roof extension, we use symmetry properties of certain multiparty quantum states and simplify the evaluation of GGM for these classes of mixed states, as prescribed in Refs. \cite{Vollbrecht,Vollbrecht2,mixed_tangle} (cf. \cite{symmetry_ent}). Exploiting such symmetries, we are able to compute the GGM of different paradigmatic classes of mixed states having different ranks. In particular, we first present the exact value of GGM for certain classes of rank 2 and rank 3 mixed states with arbitrary number of qubits. We then compute the GGM for a specific class of states which is a mixture of Greenberger-Horne-Zeilinger (GHZ) \cite{GHZ} and all the Dicke states \cite{Dicke}, 
 having a variety of ranks. The common property that all these classes possesses is that they remain invariant under the action of same symmetric local unitary operators on each qubit. Moreover, we find the GGM of a class of tripartite states of rank 4 
which remains unaltered under different local unitaries on each party. 
Finally, we show that such symmetry properties can lead to an exact expression of GGM for a class of multiqudit states having varied ranks.

The paper is organized in the following manner. In Sec. \ref{sec:GGM_def}, we review the definition and the various properties of the
generalized geometric measure for pure states. In section \ref{sec:GGM_def_mixed}, we define GGM  for mixed states via the convex roof construction. Here, we also discuss the Terhal-Vollbrecht-Werner technique of exploiting the symmetry of a quantum state for simplifying the evaluation of a convex roof extension. The same section also contains the computation of the GGM for different classes of mixed states. We present a summary in Sec. \ref{sec:conclusion}. 

\section{Generalized geometric measure}
\label{sec:GGM_def}
A pure state is said to be genuinely multiparty entangled if it is not product in any bipartition. The generalized geometric measure (GGM) \cite{GGM}  (cf. \cite{GM}) of an $N$-party pure quantum state, $|\psi_N\rangle$, is a computable entanglement measure that can quantify genuine multiparty entanglement. It is defined as an optimized  distance of the given state from the set of all states that are not genuinely multiparty entangled. Mathematically, it is given by 
\begin{equation}
\mathcal{E}(|\psi_N\rangle)=1-\Lambda_{max}^2(|\psi_N\rangle),
\end{equation}    
where $\Lambda_{\max} (|\psi_N\rangle ) = \max | \langle \chi|\psi_N\rangle |$, with the maximization being over all  $|\chi\rangle$ that are  not genuinely multiparty entangled. An equivalent form of the above equation is \cite{GGM}
\begin{equation}
\mathcal{E} (|\psi_n \rangle ) =  1 - \max \{\lambda^2_{ I : L} |  I \cup  L = \{A_1,\ldots, A_N\},  I \cap  L = \emptyset\},
\label{GGM}
\end{equation}
where \(\lambda_{I:L}\) is  the maximal Schmidt coefficient in the  bipartite split \(I: L\) of \(| \psi_N \rangle\).

Let us enumerate some properties of the GGM which establish it as a bona fide measure of genuine multiparty entanglement \cite{GGM}. $\mathcal{E} (|\psi_N \rangle )\geq0$, for all $|\psi_N \rangle$, $\mathcal{E} (|\psi_N \rangle )=0$ iff $|\psi_N \rangle$ is not genuinely multiparty entangled, and $\mathcal{E} (|\psi_N \rangle )$ is nonincreasing under local quantum operations at the $N$ parties and classical communication between them.

\section{GGM for mixed states}\label{sec:GGM_def_mixed}
We can now define the GGM of a general mixed quantum state, in terms of the convex roof construction. 
For an arbitrary $N$-party mixed state, $\rho_N$, the GGM can be defined as
\begin{equation}
\mathcal{G}(\rho_N)=\min_{\{p_i, |\psi^i_N\rangle\}}\sum_i p_i\mathcal{E}(|\psi_N^i \rangle), 
\end{equation} 
where the minimization is over all pure state decompositions of $\rho_N$ i.e., $\rho_N=\sum_i p_i|\psi_N^i \rangle\langle\psi_N^i|$. It is difficult to find the optimal decomposition  and the computation of GGM is
in general impossible even for moderate-sized systems.  However, the  
situation is different if the mixed quantum state under consideration possesses some symmetry \cite{symmetry_ent,mixed_tangle, Vollbrecht, Wei_Goldbart}. In Ref. \cite{Vollbrecht}, Vollbrecht and Werner have provided a general method to compute an entanglement measure, defined via the convex roof extension, of a class of mixed states which are invariant, on average, under a group of local unitaries. Below we briefly outline the same. 
Suppose $\rho'_N= (U_1\otimes U_2\otimes \ldots \otimes U_N)\rho_N(U_1^{\dagger}\otimes U_2^{\dagger}\otimes \ldots \otimes U_N^{\dagger})$, where $U_i$ are the local unitary operators, acting on Hilbert spaces $ H_i$. 
The GGM of $\rho_N$ and $\rho'_N$ are the same. If it happens that $\rho_N=\rho'_N$, then $(U_1\otimes U_2\otimes \ldots \otimes U_N)$ is called a local symmetry of $\rho_N$. Let $G$ be a group of unitary operators $U=(U_1\otimes U_2\otimes \ldots \otimes U_N)$ and ${\bf P}$ be a twirl operator, such that, 
$ A \overset{\bf P}{\longrightarrow} \int dU~UAU^\dagger \equiv {\bf P}(A)$,
where the integral is carried out Haar uniformly. 
In case of a  mixed state $\rho_N$, if there exist a twirl operator ${\bf P}$ such that $ {\bf P}(\rho_N)=\rho_N$, then the entanglement, ${\cal G}(\rho_N)$, can be obtained from a pure $|\psi\rangle$ which satisfies 
\begin{equation}\label{Eq:Projector_def}
{\bf P}(|\psi\rangle\langle\psi|)=\rho.
\end{equation} 
 In principle, one can have a set of pure states,  $\{|\psi\rangle\} = M_{\rho_N}$, which satisfies Eq. (\ref{Eq:Projector_def}), and it is sufficient to perform the optimization over this set. 
A further step is needed where we convexify the optimized quantity over the parameters in $\rho_N$, if it is not already convex.

We now show that this method can be utilized to evaluate the GGM for several classes of multiparty states with arbitrary number of parties having certain symmetries. We present these classes according to their ranks. 

\begin{figure}[t]
\begin{center}
  \includegraphics[scale=0.58]{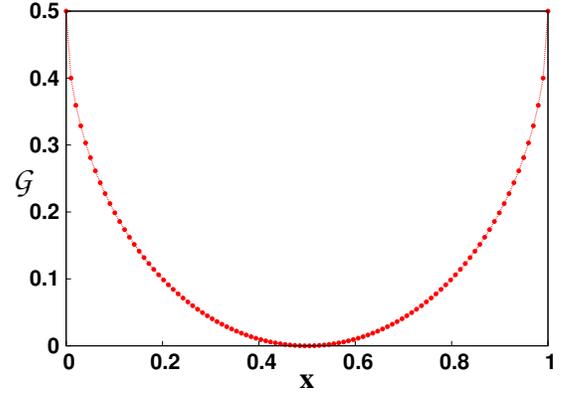}
\end{center}
\caption{(Color online.) GGM of $\rho^2_N(x)=x |\psi_N\rangle \langle\psi_N| + (1-x) |\psi_N^\perp\rangle\langle \psi_N^\perp|$ against x. All the quantities are dimensionless.} 
\label{fig:sigmaz_invariant}
\end{figure}

\subsection{Classes of rank 2 multiqubit states}\label{sebsec:rank_2}
The rank 2 mixed state, which we are now going to consider is a  mixture of two orthogonal $N$-party pure states,  given by
\begin{eqnarray}
 \rho^2_N(x)=x |\psi_N\rangle \langle\psi_N| + (1-x) |\psi_N^\perp\rangle\langle \psi_N^\perp|,
 \end{eqnarray} 
 where the subscript and superscript of $\rho$ represent  the  number of qubits and rank respectively. Here, $|\psi_N\rangle$ and $|\psi_N^\perp\rangle$ lie in two orthogonal mutually complementary subspaces of the N-party Hilbert space ${\cal H}^{\otimes N}$.  $|\psi_N\rangle =\sum_{i=0}^{\lfloor\frac{N}{2}\rfloor} a_i|D^{2i}_g\rangle$, with
\begin{eqnarray}\label{Eq:Dicke_state_def}
|D^{k}_g\rangle =\sum_{j=1}^{\binom Nk}b_{kj}|\underbrace{00...0}_{N-k}\underbrace{11..1}_{k}\rangle,
\end{eqnarray}
 where 
$|D^{k}_g\rangle$'s are the generalized Dicke states \cite{Dicke} with $k$ number of excitations i.e. they are the general
superpositions of pure states with 
all permutations of $(N-k)$ $|0\rangle$'s and $k$ $ |1\rangle$'s.
And
\begin{eqnarray}
|\psi_N^\perp\rangle =\sum_{i=0}^{\lfloor\frac{N}{2}\rfloor - 1}a'_i |D^{2i+1}_g\rangle.
\end{eqnarray}
We have chosen the coefficients in all pure and mixed states such that there are properly normalized. 

For $\rho^2_N(x)$, we can find a group of local unitary operators consisting of two unitaries, $U_1=I$, and $U_2=\sigma_z$, which, on average, keep $\rho^2_N(x)$ invariant.
Here, $I$ is the identity operator on the qubit Hilbert space and $\sigma_x$, $\sigma_y$, and $\sigma_z$ are the Pauli operators.
One can check that $\rho^2_N(x) = \sum_{k =1}^2 U_k^{\otimes N} |\psi^2_N(x)\rangle \langle\psi^2_N(x)|U_k^{\dagger \otimes N}$, where $|\psi^2_N(x)\rangle=\sqrt{x}|\psi_N\rangle+ e^{i\phi}\sqrt{1-x} |\psi_N^{\perp} \rangle$ is the only class of pure states that is twirled to $\rho^2_N (x)$ by  applying the twirl operator corresponding to those unitaries.
 Hence, by following the recipe in \cite{Vollbrecht}, we can calculate the GGM of $\rho^2_N(x)$. Since it involves several parameters, for illustration, we choose fully symmetric states, i.e, when all the coefficients of $|\psi_N\rangle$ and $|\psi_N^\perp\rangle$  are equal. The GGM of  $\rho^2_N(x, sym)$ is the convex hull of the GGM  of the pure states $|\psi_N^2(x,sym)\rangle = \sqrt{x} |\psi_N\rangle + \sqrt{1-x}e^{i\phi_{min}} |\psi_N^\perp\rangle$. Here the phase, $\phi_{min}$, gives the minimum GGM among all the GGM with different $\phi$ values. We then find that GGM reaches its 
minimum for  $\phi_{min} = 0$. Therefore, the GGM of $\rho^2_N(x, sym)$ is given by
\begin{eqnarray}
\label{Eq:GGMGHZ}
\mathcal{G}(\rho^2_N(x,sym)) = ~\frac{1}{2}(1-2\sqrt{x}\sqrt{1-x}),
\end{eqnarray}
since the right hand side is already convex as depicted in Fig.~\ref{fig:sigmaz_invariant}. An important point to note here that the GGM of $\rho^2_N(x, sym)$, given in Eq. (\ref{Eq:GGMGHZ}), is independent of number of parties, $N$.  

%
  
\begin{figure}[t]
\begin{center}
  \includegraphics[scale=0.75]{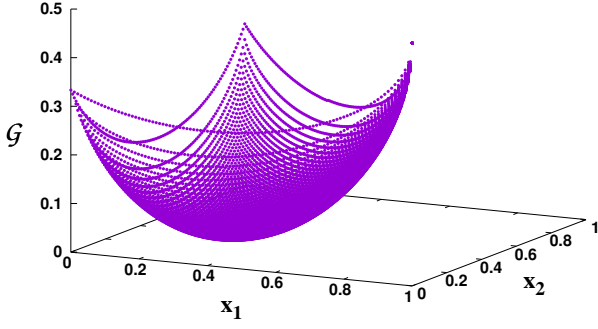}
\end{center}
\caption{(Color online.) A plot of the GGM of $\rho_3^3(x_1,x_2)=x_1~|GHZ_3^+\rangle \langle GHZ_3^+ |+ x_2~|D^1\rangle \langle D^1|+ (1-x_1-x_2)|D^2\rangle \langle D^2|$ with the state parameters $x_1$ and $x_2$. All the axes are dimensionless.}
\label{fig:ghzww'}
\end{figure}  
 
 \subsection{Classes of rank 3 multiqubit states}\label{subsec:rank_3}
 We now calculate the GGM for different classes of mixed states, of rank 3.

\subsubsection[short]{Case 1}

Let us now consider a three-qubit rank 3 mixed state, $\rho_3^3(x_1,x_2)$ \cite{Wei_Goldbart}, which is a mixture of known $|GHZ_3^+\rangle$, $|D^1\rangle$, and $|D^2\rangle$. Here, $|GHZ_3^+\rangle=\frac{1}{\sqrt{2}}(\vert 000 \rangle+\vert 111\rangle )$ \citep{GHZ}, and  
$|D^1\rangle$ and $|D^2\rangle$ are given by $|D_g^1\rangle$ and $|D_g^2\rangle$ of Eq. (\ref{Eq:Dicke_state_def}) respectively, with $b_{kj}=\frac{1}{\sqrt{3 }}$ for all $j$.
It reads as 
\begin{eqnarray}\label{Eq:rank_3_mixed1}
 \rho_3^3(x_1,x_2)& = x_1~|GHZ_3^+\rangle \langle GHZ_3^+ |+ x_2~|D^1\rangle \langle D^1| \nonumber
 \\ &+ (1-x_1-x_2)|D^2\rangle \langle D^2|. 
\end{eqnarray}
 Note that $|D^1\rangle$ is the well-known W-state \cite{zhgstate}.
The mixture is invariant under local unitaries given by $U_1 = I$, $U_2 = \left( \begin{matrix} 1 & 0\\ 0 & e^{\frac{2\pi i}{3}} \end{matrix} \right)$, and $U_3 = \left( \begin{matrix} 1 & 0\\ 0 & e^{\frac{- 2\pi i}{3}} \end{matrix} \right)$, when they act on each qubit \cite{Wei_Goldbart}. The corresponding pure state which after local unitary transformations, leads to $\rho_3^3(x_1,x_2)$, can be written as   
\begin{eqnarray}\label{Eq:GHZ_D1_D2}
|\psi_3^3(x_1,x_2)\rangle &=&  \sqrt{x_1}|GHZ^+\rangle+\sqrt{x_2}e^{i\phi_1}|D^1\rangle \nonumber \\
&& \hspace{0.6in} + \sqrt{1-x_1-x_2}e^{i\phi_2}|D^2\rangle.
\end{eqnarray}
The minimum of GGM among $\{\phi_1,\phi_2\}$  is again obtained when $\phi_1 = \phi_2 = 0$. By computing the Hessian matrix, we find both analytically and numerically that the GGM of $|\psi_3^3(x_1,x_2)\rangle$ is convex with respect to $x_1$ and $x_2$. Therefore, the GGM of $\rho_3^3(x_1,x_2)$ is given by
\begin{eqnarray}
{\cal{G}}\left(\rho^3_3(x_1,x_2)\right)&=&\frac{1}{6}\bigg(3-\bigg\{1-5x_1^2-12x_2 (x_2-1) + \nonumber \\ && \hspace{-0.5in} 8\sqrt{6x_1x_2}\left(1+\sqrt{x_2(1-x_1-x_2)}-x_1-x_2\right)+ \nonumber \\
 && \hspace{-0.5in}  4x_1 \left(1+3\sqrt{x_2(1-x_1-x_2)} -3x_2\right)
 \bigg\}^\frac{1}{2}\bigg), 
\end{eqnarray}
and is depicted in Fig. \ref{fig:ghzww'}.
%
 
 \begin{figure}[t]
\begin{center}
  \includegraphics[scale=0.75]{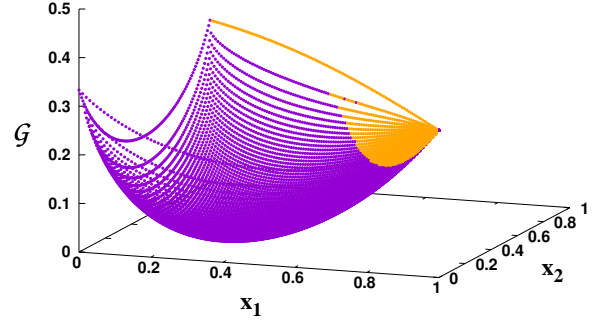}
\end{center}
\caption{(Color online.) Plot corresponds to GGM of $|\psi_3^{3, g}\rangle $ vs. the mixing parameters $x_1$ and $x_2$. Here, $\alpha = 0.55$ for the $|gGHZ_3\rangle$ state. 
Both convex and nonconvex regions are seen. The convex part corresponds to the GGM of 
$\rho_3^{3, g}(x_1, x_2)$. All quantities are dimensionless. 
} 
\label{fig:gghzww'}
\end{figure}

\subsubsection[short]{Case 2}\label{subsubsec:rank_3_gener}
Let us now move to a more general state while keeping the rank fixed. 
Precisely,
we consider a class of mixed states of the form
\begin{eqnarray}
\rho^{3,g}_3(x_1,x_2) &=& x_1|gGHZ_3\rangle\langle gGHZ_3| + x_2| D_g^1\rangle\langle D_g^1| \nonumber \\ 
&& \hspace{0.5in} + (1-x_1-x_2)|D_g^2\rangle\langle D_g^2|,
\end{eqnarray}
 where $|gGHZ_3\rangle = \alpha|000\rangle+\sqrt{1-\alpha^2}|111\rangle$ is the generalized Greenberger-Horne-Zeilinger state with $0\leq\alpha\leq 1$. The set of local unitaries that keep $\rho_3^3(x_1,x_2) $ invariant, also keep the state $\rho^{3,g}_3(x_1,x_2)$ invariant, and the class of pure state that are projected to $\rho^{3,g}_3(x_1,x_2)$ is given by 
 \begin{eqnarray}\label{Eq:pure_gGHZ_WW'}
|\psi_3^{3,g}(x_1,x_2)\rangle &=& \sqrt{x_1}| gGHZ_3\rangle + e^{i\phi_1}\sqrt{x_2}| D^1_g\rangle \nonumber \\ && \hspace{0.5in}+ e^{i\phi_2}\sqrt{1-x_1-x_2}|D^2_g\rangle.\,\,\,
\end{eqnarray}
In this case, we have $\rho_3^{3,g}(x_1,x_2) = \sum_{j = 1}^3 U_j^{\otimes 3}|\psi_3^{3,g}(x_1,x_2)\rangle\langle\psi_3^{3,g}(x_1,x_2)|U_j^{\dagger \otimes 3 }$, where $\{U_j, j=1,2,3\}$ is the same as in Case 1.  



\begin{figure}[t]
\begin{center}
  \includegraphics[scale=0.6]{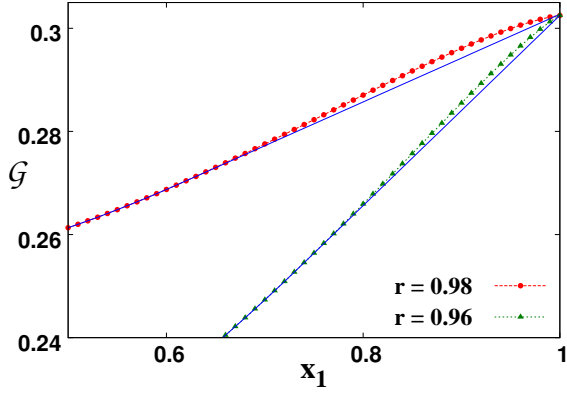}
\end{center}
\caption{(Color.) Plot corresponds to GGM of $|\psi_3^{3,g}\rangle$ vs. $x_1$, for two values of $r = \frac{x_2}{1-x_1}$. Here, $\alpha = 0.55$ for the $|gGHZ_3\rangle$ state. These are given by the dotted lines. The straight lines corresponds to the convexified quantities. All quantities are dimensionless.} 
\label{fig:gghzww'_convexify}
\end{figure}

 Numerical simulation guarantees that the minimum of ${\cal E}(|\psi_3^{3,g}(x_1,x_2)\rangle)$ occurs for $\phi_1=\phi_2=0$. However, unlike the previous cases, we find that ${\cal E}(|\psi_3^{3,g}(x_1,x_2)\rangle)$ is not convex for all values of $x_1$ and $x_2$. In particular, we plot ${\cal E}(|\psi_3^{3,g}(x_1,x_2)\rangle)$ in Fig. \ref{fig:gghzww'}, when $\alpha = 0.55$ and when the coefficients in $|D_g^1\rangle$ and $|D_g^2\rangle$ are all equal. 
 For certain regions of the parameter space, the figure is already convex, and hence the GGM of $|\psi_3^{3,g}(x_1,x_2)\rangle)$  in that region is the GGM of $\rho_3^{3,g}(x_1, x_2)$. On the other hand, 
 for the remaining regions, a convexification has to be carried out to obtain the GGM of $\rho_3^g(x_1,x_2)$.
  Specifically, ${\cal E}(|\psi_3^{3,g} (x_1,x_2)\rangle) \neq {\cal G} (\rho_3^{3,g}(x_1,x_2))$, when $x_1$ is high while $x_2$ is low. 
   To obtain the GGM in that region, the convexification is required. To illustrate the process, we introduce a new variable, $r=\frac{x_2}{1-x_1}$, and let us consider cases where $r=0.96$ and $0.98$. The convexification of the curves so generated are depicted in Fig. \ref{fig:gghzww'_convexify}.
\

\subsubsection[short]{Case 3}
 
\begin{figure}[t]
\begin{center}
  \includegraphics[scale=0.75]{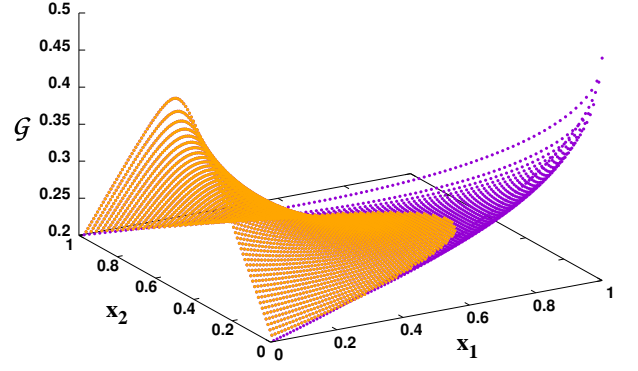}
\end{center}
\caption{(Color online.) The plot of GGM for $\rho^3_5(x_1,x_2) = x_1~|GHZ_5^+\rangle \langle GHZ_5^+ |+ x_2~|D^1\rangle \langle D^1| + (1-x_1-x_2)|D^{4}\rangle \langle D^{4}|$ against $x_1$ and $x_2$ whenever it is convex. All axes are dimensionless.} 
\label{fig:rank3_5qubit}
\end{figure}

  Let us move to a class of states which is a multiqubit generalization of $\rho_3^3(x_1,x_2)$. It is given by
  \begin{eqnarray}
  \rho^3_N(x_1,x_2) &=& x_1~|GHZ_N^+\rangle \langle GHZ_N^+ |+ x_2~|D^1\rangle \langle D^1| \nonumber\\
  && \hspace{0.3in} + (1-x_1-x_2)|D^{N-1}\rangle \langle D^{N-1}|,
  \end{eqnarray} 
 where $|GHZ_N^+\rangle = \frac{1}{\sqrt{2}}(|0\rangle^{\otimes N} + |1\rangle^{\otimes N})$,  and $|D^{N-1}\rangle$ is given by 
  $|D^{N-1}_g\rangle$ of Eq. (\ref{Eq:Dicke_state_def}) with $b_{kj} = \frac{1}{\sqrt{\binom Nk }}$.
 Again, we have $\rho^3_N(x_1,x_2) = \sum_{j = 1}^3 U_j^{\otimes N}|\psi^3_N(x_1,x_2)\rangle\langle\psi^3_N(x_1,x_2)|U_j^{\dagger \otimes N }$, where $|\psi^3_N(x_1,x_2)\rangle$ is given in Eq. (\ref{Eq:GHZ_D1_D2}) with $|D^2\rangle$ being replaced by $|D^{N-1}\rangle$, for the same set of unitaries, given in Case 1. Hence, we can compute the GGM of $|\psi^3_N(x_1,x_2)\rangle$ and check its convexity. 
 For $\phi_1 = \phi_2 = 0$ which gives the lowest GGM, 
  Fig. \ref{fig:rank3_5qubit} shows the GGM of $|\psi^3_5(x_1,x_2)\rangle$ with respect to the parameters, $x_1$ and $x_2$ with $N = 5$. From the figure, it is clear that for example the GGM of $|\psi^3_5(x_1,x_2)\rangle$  is convex for $ 0.64 \leq x_1 \leq 1.0$ and $0.0 \leq x_2 \leq 0.36$ and hence in that region, we have the GGM of $\rho^3_5(x_1,x_2)$. In the rest of the region, to obtain the GGM of $\rho^3_5(x_1,x_2)$, we have to find the convex hull of ${\cal E}(|\psi^3_5(x_1,x_2)\rangle)$.

\subsection{Higher rank multiqubit states} \label{subsec:rank_higher}
We now consider classes of mixed states with rank more than three. First, we explore a class of multiparty states which can be dealt with symmetric unitaries. In other words, this class of states remain invariant, when the same unitary acts on all the parties, i.e. $\rho_N^N = \sum_j U_j^{\otimes N} \rho_N^N U_j^{\dagger \otimes N}$.  We will then find another class of states for which symmetric unitaries do not work.   

\subsubsection{ Symmetric unitary case}
Let us now consider a class of mixed states with arbitrary number of parties, which can be obtained by generalizing $\rho_3^3(x_1,x_2)$. The state, $\rho_N^N(x_1,x_2,\ldots, x_{N-1} )$, is a mixture of generalized GHZ and all the Dicke states. It reads as 
\begin{eqnarray}
\rho_N^N(x_1,x_2,\ldots, x_{N-1} ) &=& (1 - \sum_i x_i) |gGHZ_N\rangle\langle gGHZ_N|  \nonumber \\
&& \hspace{0.6in} + \sum_{i=1}^{N-1} x_i|D_g^{i}\rangle\langle D_g^{i}|, 
\end{eqnarray}
with $|gGHZ_N\rangle = \alpha|0\rangle^{\otimes N} + \sqrt{1-\alpha^2}|1\rangle^{\otimes N}$. Rank of the above state spans the integers in $[1,N]$. One can check that 
\begin{eqnarray}
\rho_N^N(x_1,\ldots, x_{N-1}) = \sum_{j = 1}^N U_j^{\otimes N}\rho_N^N(x_1,\ldots, x_{N-1})U_j^{\dagger \otimes N}, \nonumber \\
\end{eqnarray}
where the set of local unitaries, $\{U_j\}_{j =1}^N$ consists of $I$ and $\left( \begin{matrix} 1 & 0\\ 0 & e^{\frac{2\pi i j }{N}} \end{matrix} \right)$ with $j = 1,\ldots, (N-1)$. We have to now show that 
\begin{eqnarray}
\label{Eq:unitaryN}
\rho_N^N(x_1,x_2,\ldots, x_{N-1}) &=& \sum_j U_j^{\otimes N} |\psi_N^N(x_1,\ldots,x_{N-1} )\rangle \nonumber \\
&& \langle \psi_N^N(x_1,\ldots,x_{N-1}| U_j^{\dagger \otimes N}, 
\end{eqnarray}
where $|\psi_N^N(x_1,\ldots,x_{N-1} )\rangle = \sqrt{1 - \sum_i x_i} |gGHZ_N\rangle + \sum_{i=1}^{N-1} \sqrt{x_i}|D_g^{i}\rangle$. To prove this, the we note the actions of local unitaries on each off-diagonal terms which e.g. are given by
\begin{equation}
U_j^{\otimes N} |D_g^q\rangle\langle D_g^r|U_j^{\dagger\otimes N}  = e^{\frac{2\pi i(q-r)}{N}}|D_g^q\rangle\langle D_g^r|. 
\end{equation}  
We use the identity $ \sum_j e^{\frac{2\pi i(q-r)}{N}} = \delta_{qr}$ in the analysis. Similarly,
\begin{equation}
\sum_j U_j^{\otimes N} |D_g^q\rangle\langle gGHZ_N |U_j^{\dagger\otimes N}  = e^{\frac{2\pi iq}{N}}|D_g^q\rangle\langle gGHZ_N| = 0.
\end{equation} 
All off-diagonal terms therefore vanish. We can now calculate the GGM of $|\psi_N^N(x_1,\ldots,x_{N-1} )\rangle$ and check whether ${\cal E}(|\psi_N^N(x_1,\ldots,x_{N-1} )\rangle)$ is convex or not. If it is convex, then ${\cal E}(|\psi_N^N(x_1,\ldots,x_{N-1} )\rangle) = {\cal G}(\rho_N^N(x_1,\ldots, x_{N-1}))$. Otherwise, we have to perform convexification to obtain the exact value of  ${\cal G}(\rho_N^N(x_1,\ldots, x_{N-1}))$. To illustrate this example, we consider a five-qubit state which is of the form  
\begin{eqnarray}
&& \rho_5^5 = x_1 |GHZ_5^+\rangle\langle GHZ_5^+| + \frac{x_2}{2} (|D^1\rangle \langle D^1| + |D^2\rangle \langle D^2|)\nonumber 
 \\
 && \hspace{0.9in} + \frac{1 - x_1 - x_2}{2}(|D^3\rangle \langle D^3| + |D^4\rangle \langle D^4|).\nonumber \\
\end{eqnarray}

\begin{figure}[t]
\begin{center}
  \includegraphics[scale=0.75]{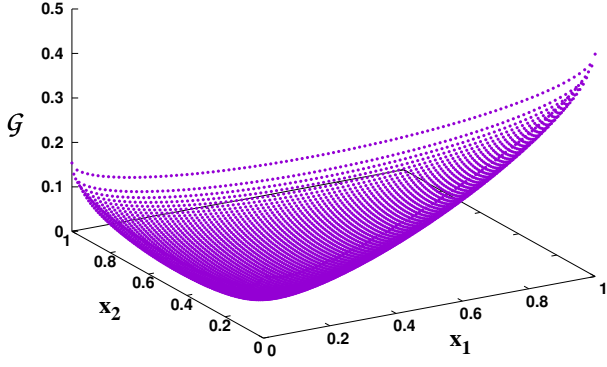}
\end{center}
\caption{(Color online.) GGM of $\rho^5_5= x_1 P [GHZ_5^+] + \frac{x_2}{2} (P [D^1]+P [D^2])+\frac{1-x_1-x_2}{2}(P [D^3]+P [D^4])$. All axes are dimensionless.} 
\label{fig:rank5_5qubit}
\end{figure}

Following the aforementioned prescription, we compute ${\cal E}(|\psi_5^5(x_1,x_2)\rangle)$ with 
\begin{eqnarray}
&&|\psi_5^5(x_1,x_2 )\rangle =  \sqrt{x_1} |GHZ_5^+\rangle  + \sqrt{\frac{x_2}{2}} \sum_{k=1}^{2}  e^{i \phi_k}|D^k\rangle \nonumber 
 \\
 && \hspace{1in}+ \sqrt{\frac{1-x_1-x_2}{2}} \sum_{k=3}^{4}  e^{i \phi_k}|D^k\rangle.
\end{eqnarray}
For $\phi_k = 0, k = 1,\ldots,4$ which gives the infimum of GGM, ${\cal E}(|\psi_5^5(x_1,x_2 )\rangle)$ is plotted with $x_1$ and $x_2$ in Fig. \ref{fig:rank5_5qubit}. By using the Hessian technique, we find that it is convex for the entire range of $x_1$ and $x_2$. Therefore, ${\cal G}(\rho_5^5)$ is obtained for all $x_1$ and $x_2$ and is given by
\begin{eqnarray}
{\cal G}(\rho_5^5) = \frac{1}{2}\Bigg( 1- \bigg(1-4 \bigg\{ \frac{2x_1 + 4x_2+3}{10} ~~ \frac{7-2x_1 -4x_2}{10} - 
\nonumber \\ \Big( \sqrt{\frac{x_1x_2}{20}} + \sqrt{\frac{x_1(1-x_1 -x_2)}{20}} + \frac{2x_2}{5\sqrt{2}} + \frac{2(1-x_1-x_2)}{5\sqrt{2}} \nonumber \\
+ \frac{3}{10}\sqrt{x_2(1-x_1-x_2)}\Big)^2 \bigg\} \bigg)^{\frac{1}{2}}\Bigg). \hspace{0.3in}
\end{eqnarray}
Comparing Figs. \ref{fig:rank3_5qubit} and \ref{fig:rank5_5qubit} with the situations obtained before, it seems that higher rank states, for a fixed total number of qubits of the entire systems, have a greater affinity for being convex, when their GGMs are considered.


\subsubsection{ Asymmetric  unitary case}
\begin{figure}[t]
\begin{center}
  \includegraphics[scale=0.28]{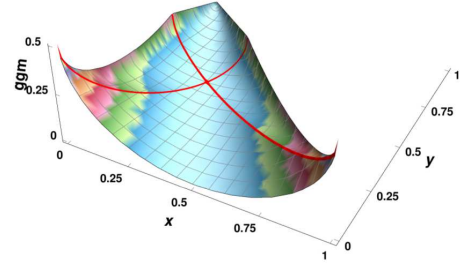}
\end{center}
\caption{(Color online.) Plot of GGM of $\rho^4_3$ with respect to the parameters, $x$ and $y$. The GGM of the corresponding unique  pure state, $|\psi^4_3(x, y)\rangle = \sqrt{x} |\zeta_1\rangle - i\sqrt{y/2} ( |\zeta_2\rangle -  |\zeta_3\rangle) + \sqrt{1-x-y}  |\zeta_4\rangle$ has a kink along the lines shown on the surface, in the plot. The GGM of the pure state is non-convex around these lines, and hence convexifications are required thereat. } 
\label{fig:rank4_3qubit}
\end{figure}

Until now, we have considered the states which remain unaltered under local symmetric unitaries of the form $U_i^{\otimes ^N}$. 

Let us now illustrate a class of three-qubit mixed states which remains unchanged under the local unitaries of the form $U_i\otimes U_j\otimes U_k$. The class of mixed state having rank 4, reads 
\begin{eqnarray}
\rho^4_3 = \sum_i x_i |\zeta_i\rangle \langle \zeta_i|, 
\end{eqnarray}
where 

\begin{eqnarray}\nonumber
 \ket{\zeta_1} &=& \frac{1}{2}( |001 \rangle + |010\rangle - |100\rangle + |111\rangle), \nonumber\\
 \ket{\zeta_2} &=& \frac{1}{2}(-i |000 \rangle - i |011\rangle + |100\rangle + |111\rangle),\nonumber\\ 
 \ket{\zeta_3} &=& \frac{1}{2}( i |000 \rangle + i|011\rangle + |100\rangle + |111\rangle), \nonumber\\ 
\text{and\,\,} \ket{\zeta_4} &=& \frac{1}{2}( |001 \rangle + |010\rangle + |100\rangle - |111\rangle). \nonumber 
\end{eqnarray}

It is invariant under $\{U_i, i =1, \ldots 4\}$, which are given by 
\begin{eqnarray} \nonumber
U_1 &=& I \otimes I \otimes I, \\ \nonumber
U_2 &=& i \sigma_y  \otimes H' \otimes H', \\ \nonumber
U_3 &=&  I  \otimes \sigma_y \otimes \sigma_y,  \\ \nonumber
\text{and\,\,}  U_4 &=& - i \sigma_y  \otimes H'^T \otimes H'^T,  \nonumber
\end{eqnarray}
with $H' = \frac{1}{\sqrt{2}}\left( \begin{matrix} 1 & 1\\ -1 & 1 \end{matrix} \right)$. Note that these unitaries form a closed group. 
The only pure states that are twirled to the above mixed states are of the form $\ket{\psi^4_3}= \sum_i \sqrt{x_i}e^{i\phi_i} |\zeta_i\rangle $. We compute the GGM of $\ket{\psi^4_3}$ and minimize it over $\phi_i$'s. The GGM of $\rho_4^3$ is given by the minimum of the ${\cal E}(\ket{\psi^4_3})$ for different values of \(\phi_i\)s provided the quantity is convex itself. 

To visualize its GGM, let us consider, $x_2 = x_3 = \frac{y}{2}$, i.e. the state is of the form
\begin{eqnarray}
\rho_4^3 &=& x \ket{\zeta_1} \bra{\zeta_1} + \frac{y}{2}(\ket{\zeta_2}\bra{\zeta_2} + \ket{\zeta_3}\bra{\zeta_3}) \nonumber \\ && \hspace{1.2in}+ (1-x-y)\ket{\zeta_4}\bra{\zeta_4}.
\end{eqnarray}
In this case, we find that the minimum GGM of  $|\psi^4_3(x,y)\rangle$ for different values of \(\phi_i\)'s is obtained when $\phi_1 = -~\phi_2= - \frac{\pi}{2}$ and $\phi_3 = 0$. We find the GGM of $\rho_3^4 (x, y)$ by convexifying the GGM of 
$|\psi^4_3(x,y)\rangle)$.

\vspace{0.2in}

\subsection{Cases of multiqudit states} \label{subsec:multiqudit_multiparty}

In the previous sections, we have evaluated the GGM of certain multiqubit systems. 
We will now show that a similar method can be extended to obtained the analytical expression of GGM of multiqudit mixed states. 
 Specifically, we consider an $N$-qudit mixed state of rank $d$, in the Hilbert space $\mathcal{H}_d^{\otimes N}$, of the 
form 
\begin{equation}
\label{d-mix}
 \rho_{N,d}^d = \sum_{k=1}^{d}p_k|\Psi\rangle_k\langle\Psi|_k,
\end{equation}
where $| \Psi\rangle_k = \sum_{\{j\}}q_{j_1j_2...j_N}| j_1j_2...j_N\rangle^{(k)}$ and  $(\sum_{m}j_m)(\text{mod}~d)=k$.
 Our aim is to evaluate the GGM of the state $\rho_{N,d}^d$. Therefore, like previous cases, we construct a twirling operator, consisting of unitary operators $Z_d$ which are $d$-dimensional, non-hermitian generalization of the $\sigma_z$ and given by
\begin{eqnarray}
\label{d-unitary}
 Z_d= \sum_{j=0}^{d-1}e^{\frac{2\pi ij}{d}}|j\rangle\langle j|.
\end{eqnarray}
Here, each of the unitary operators act locally and symmetrically on $\rho_{N,d}^d$ as $Z_d^{\otimes N}$. 
Note that the set $\bigg\{ I_d, Z_d^{\otimes N}, \Big(Z_d^{\otimes N}\Big)^2, .., \Big(Z_d^{\otimes N}\Big)^{d-1}\bigg\}$ forms a group and the corresponding twirling operator keeps $\rho_{N,d}^d$ invariant. Now, we have to find the set of all pure states $|\Psi\rangle_{N,d}^d$ that are projected 
to $\rho_{N,d}^d$ under the action of the aforementioned twirling operator.
It can be easily checked that $|\Psi\rangle_{N,d}^d = \sum_{k=1}^de^{i\phi_k}|\Psi\rangle_k$ are the only class of pure states that are mapped to  $\rho_{N,d}^d$ under the twirling operator, i.e., $\sum_{q=0}^{d-1}\Big(Z_d^{\otimes N}\Big)^q|\Psi\rangle_{N,d}^d\langle \Psi|_{N,d}^d \Big(Z_d^{\dagger \otimes N}\Big)^q=\rho_{N,d}^d$. In this case also, the minimum of the GGM's of $|\Psi\rangle_{N,d}^d$ over the phases $\{\phi_k\}$
gives the GGM of $\rho_{N,d}^d$ provided the minimum GGM is already a convex function of the state parameters. Otherwise one has to convexify the function to obtain the GGM of $\rho_{N,d}^d$.

Until now, we have considered systems with the same dimensions of the local Hilbert spaces. However, this formalism can be further extended where the local Hilbert spaces' dimensions are not equal, i.e., for quantum systems belonging in 
$\mathcal{H}_{d_1}\otimes \mathcal{H}_{d_2}\otimes \ldots \otimes \mathcal{H}_{d_N}$, with $d_1\neq d_2\neq ...d_N$. In that case, we have two different scenarios. 
Firstly, $a_1d_1=a_2d_2=...=d_N$, where $\{a_i\}_{i=1}^{N-1}\in\mathcal{I^+}$. Without loss of generality, $d_N$ is taken  to be the largest dimension and the corresponding unitaries are of the form 
$Z_{d_1}\otimes Z_{d_2}..\otimes Z_{d_N}$ with its subsequent powers upto $d_N-1$, such that the composite unitary matrices form a group.
Evidently, the case of equal dimensions is a special case of this. 
Thus, the pure states over which we have to perform the minimization still have the same form, with a slightly different version of the condition given by
$\sum_{m}j_m(\text{mod}~d_N) =k$.
The second one is the situation when all the dimensions are prime to each other, and in this case, we have to take unitaries upto the power of $\big(d_1d_2...d_N\big)-1$, where the form of pure states remain the same, with 
the modified condition, $\sum_{m}j_m\big(\text{mod}~d_1d_2...d_N\big) =k$. Therefore, in general, we have to take the maximum power of the unitaries which is the lowest common multiple of $ d_1,d_2,...,d_N$ to apply the similar prescription. In the next paragraph, we illustrate this with an example.

For simplicity, we consider the following three-qutrit state, $\rho_{3,3}^3 = \sum_{k=0}^2x_k|\Psi\rangle_k\langle\Psi\vert_k$, where 
$|\Psi\rangle_k = \sum_{j}q_{j_1j_2j_3}| j_1j_2j_3\rangle^{(k)}$ and $j_1+j_2+j_3(\text{mod}~ 3)=k$. The exact form of the pure states $\{|\Psi_k\rangle\}_{k=0}^2$ reads as
\begin{eqnarray}\nonumber
\label{3-qutrit-pure}
 \ket{\Psi_0} &=& \frac{1}{3}( \sum_{i=0}^2|iii \rangle + \sum_{perm}|012\rangle ), \nonumber\\
 \ket{\Psi_1} &=& \frac{1}{3}( \sum_{perm}|001 \rangle +\sum_{perm} |022\rangle + \sum_{perm}|112\rangle),\nonumber\\ 
\hspace{-0.5in}\text{and} ~~ \ket{\Psi_2} &=& \frac{1}{3}( \sum_{perm} |011 \rangle \sum_{perm}|002 \rangle+ \sum_{perm}|122\rangle). 
\end{eqnarray}
For this case, the unitaries which construct the twirling operators are given as $\{ I_3, Z_3, Z_3^2\}$. 
Note that the unitaries of the form $ Z_3^i\otimes Z_3^i\otimes Z_3^i$ form a group
for $i$ ranging from $0$ to $2$ and $\rho_{3,3}^3$ is evidently invariant under the corresponding twirling operator. 
The pure state that is mapped to $\rho_{3,3}^3$ under the action of the aforesaid twirling operator is of the form $|\Psi^3_{3,3}\rangle =\sqrt{x_1}|\Psi\rangle_1+e^{i\phi_2}\sqrt{x_2}|\Psi\rangle_2+e^{i\phi_3}\sqrt{1-x_1-x_2}|\Psi\rangle_3 $. 
\begin{figure}
\begin{center}
  \includegraphics[scale=0.76]{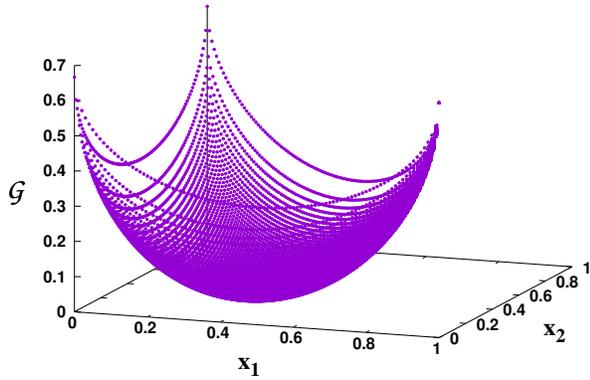}
\end{center}
\caption{(Color online.) Plot of GGM of $\rho^3_{3,3}$ against $x_1$ and $x_2$. The GGM of the corresponding unique  pure state, 
$|\Psi^3_{3,3}\rangle =\sqrt{x_1}|\Psi\rangle_1+e^{i\phi_2}\sqrt{x_2}|\Psi\rangle_2+e^{i\phi_3}\sqrt{1-x_1-x_2}|\Psi\rangle_3 $ is plotted with $\phi_1= \phi_2 = 0$. 
The GGM of the pure state is convex everywhere, as evident from this plot and hence ${\cal E}(|\Psi^3_{3,3}\rangle) = {\cal G}(\rho_{3,3}^3)$. } 
\label{fig:rank3_3qubit3dimension}
\end{figure}
It can be easily found that  minimum GGM of $|\Psi^3_{3,3}\rangle$ is obtained for $\phi_2=\phi_3=0$ and it is a convex function of the parameters $x_1$ and $x_2$. Hence, the GGM of $\rho^3_{3,3}$ is given by
\begin{eqnarray}\nonumber 
\label{ggm-rho-3-3}
{\cal G}(\rho^3_{3,3})=&\frac{2}{3}\{1-\sqrt{x_1x_2}-\sqrt{x_1\{1-x_1-x_2\}}\\\nonumber 
 &-\sqrt{x_2\{1-x_1-x_2\}}\}.
\end{eqnarray}
${\cal G}(\rho^3_{3,3})$ is depicted in Fig. \ref{fig:rank3_3qubit3dimension} and the convexity of the function can be visualized from the same.
\vspace{0.2 in}

\section{Conclusion}
\label{sec:conclusion}
Computing entanglement of an arbitrary mixed state is a formidable task. The entanglement of mixed states is generally defined by constructing the convex roof over all possible pure states which  is practically impossible to compute in most of the cases. Although there exists a few bipartite measures which can be obtained for arbitrary states, the evaluation of entanglement for a mixed state in multiparty domain is still a challenging task. In this paper, we have computed a genuine multiparty entanglement measure known as generalized geometric measure of some classes of mixed states with arbitrary number of parties and dimensions by using certain symmetries.
 We evaluate the measure for several classes of multiqubit and multiqudit states having different ranks. The method, we exploited, uses a pure state that contains the same amount of entanglement as the given mixed state, and leads to the mixed state by action of a certain twirling operation.
 
\textbf{Note added:} The present work is based on a poster presentation \cite{Bag_Mis_poster} at the International Workshop on Quantum Information (IWQI-2012), Harish-Chandra Research Institute, Allahabad, India.   We thank J. Solomon Ivan for pointing out during a discussion over the poster that the same method as followed here can be used to evaluate the GGM for an arbitrary
mixture of $|GHZ^+_N\rangle$ and  $|GHZ^-_N\rangle$, where
 $|GHZ^\pm_N\rangle=\frac{1}{\sqrt{2}}(|0\rangle^{\otimes N} \pm |1\rangle^{\otimes N})$. We thank Otfried G{\"u}hne for informing us about their independent work on evaluating multipartite entanglement \cite{Otfried}, by a method that is different from the one followed in the
present work.
 

%


\begin{thebibliography}{99}
 
 
\bibitem{HHHHRMP} R. Horodecki, P. Horodecki, M. Horodecki, and K. Horodecki,
Rev. Mod. Phys. {\bf 81}, 865 (2009).

\bibitem{M_Born} M. Born, \emph{Natural philosophy of cause and chance}, (Oxford University Press, London (1949)), p. 123.

\bibitem{EPR}A. Einstein, B. Podolsky, and N. Rosen, Phys. Rev. {\bf 47}, 777 (1935).

\bibitem{BennetDC} C. H. Bennett and S. J. Wiesner, Phys. Rev. Lett. {\bf 69}, 2881 (1992).



\bibitem{TeleRef} C. H. Bennett, G. Brassard, C. Cr\'epeau, R. Jozsa, A. Peres, and W.K. Wootters, Phys. Rev. Lett. {\bf 70}, 1895 (1993).

\bibitem{Crypto} A. Ekert, Phys. Rev. Lett. {\bf 67}, 661 (1991);
N. Gisin, G. Ribordy, W. Tittel, and H. Zbinden, Rev. Mod. Phys. {\bf 74}, 145 (2002).
 


%
%
%
%

\bibitem{state_tomo} M. Beck,
Phys. Rev. Lett. {\bf 84}, 5748 (2000);
R. T. Thew, K. Nemoto, A. G. White, and W. J. Munro
Phys. Rev. A {\bf 66}, 012303 (2002);
M. Christandl and R. Renner,
Phys. Rev. Lett. {\bf 109}, 120403 (2012).

\bibitem{Q_metrology} V. Giovannetti, S. Lloyd, and L. Maccone,
Phys. Rev. Lett. {\bf 96}, 010401 (2006); 
V. Giovannetti, S. Lloyd and L. Maccone, Nat. Photonics {\bf 5}, 222 (2011).

\bibitem{Phase_transit}S. Sachdev, Quantum Phase Transistions (Cambridge
University Press, Cambridge, 2011); M. Lewenstein, A. Sanpera, V. Ahufinger, B. Damski, A. Sen(De), and U. Sen, Adv. Phys. {\bf 56}, 243 (2007); L. Amico, R. Fazio, A. Osterloh, and V. Vedral, Rev. Mod. Phys. {\bf 80}, 517 (2008).

\bibitem{Lab_exp} J. M. Raimond, M. Brune, and S. Haroche, Rev. Mod. Phys. {\bf 73}, 565 (2001); D. Leibfried, R. Blatt, C. Monroe, and D. Wineland
Rev. Mod. Phys. {\bf 75}, 281 (2003); I. Bloch, J. Dalibard, and W. Zwerger
Rev. Mod. Phys. {\bf 80}, 885 (2008); R. Blatt, and D. Wineland, Nature {\bf 453}, 1008 (2008); J. W. Pan, Z. B. Chen, C.Y. Lu, H. Weinfurter, A. Zeilinger, and M. \.Zukowski, Rev. Mod. Phys. {\bf 84}, 777 (2012).

\bibitem{Ent_entropy}  C. H. Bennett, H. J. Bernstein, S. Popescu, and B. Schumacher, Phys. Rev. A {\bf 53}, 2046 (1996).
\bibitem{concEof} S. Hill and W. K. Wootters, Phys. Rev. Lett. {\bf 78}, 5022 (1997); 
	W. K. Wootters, {\em ibid.} {\bf 80}, 2245 (1998).	
	
\bibitem{distillable}  C. H. Bennett, D.P. DiVincenzo, J. A. Smolin, and W. K. Wootters, Phys. Rev. A {\bf 54}, 3824 (1996);
E. M. Rains, {\em ibid.}  {\bf 60}, 173 (1999); {\em ibid.} {\bf 60}, 179 (1999); 
P. M. Hayden, M. Horodecki, and B. M. Terhal, J. Phys. A: Math. Gen. {\bf 34}, 6891 (2001).
\bibitem{LN} A. Peres, Phys. Rev. Lett. {\bf 77}, 1413 (1996); M. Horodecki,
P. Horodecki, and R. Horodecki, Phys. Lett. A {\bf 223}, 1 (1996);
J. Lee, M. S. Kim, Y. J. Park, and S. Lee, J. Mod. Opt. {\bf 47}, 2151 (2000); 
G. Vidal and R.F. Werner, Phys. Rev. A {\bf 65}, 032314 (2002);  
M. B. Plenio, Phys. Rev. Lett. {\bf 95}, 090503 (2005).

\bibitem{Ent_quanti} G. Vidal, J. Mod. Opt. {\bf 47}, 355 (2000);
M. A. Nielsen, Phys. Rev. Lett. {\bf 83} , 436 (2002);  D. Jonathan and M. B. Plenio
Phys. Rev. Lett. {\bf 83}, 1455 (1999).

\bibitem{GM}  A. Shimony, Ann. N.Y. Acad. Sci. \textbf{755}, 675 (1995); 
H. Barnum and N. Linden, J. Phys. A \textbf{34}, 6787 (2001);
 M. B. Plenio and V. Vedral, J. Phys. A {\bf 34}, 6997 (2001);
D. A. Meyer and N. R. Wallach, J. Math. Phys. {\bf 43}, 4273 (2002);
A. Osterloh and J. Siewert, Phys. Rev. A {\bf 72}, 012337 (2005); A. Osterloh and J. Siewert, Int. J. Quant. Inf. {\bf 4}, 531 (2006); R. Or\'us, Phys. Rev. Lett.
{\bf 100}, 130502 (2008); R. Or\'us, S. Dusuel, and J. Vidal, \emph{ibid.}
{\bf 101}, 025701 (2008); R. Or\'us, Phys. Rev. A {\bf 78}, 062332 (2008); M. Balsone, F. Dell’Anno, S. De Siena, and F. Illuminatti, Phys. Rev. A
{\bf 77}, 062304 (2008); D. Z. Djokovi\'c and A. Osterloh, J. Math. Phys.
{\bf 50}, 033509 (2009); Q.-Q. Shi, R. Or\'us, J. O. Fjærestad, and H.-Q. Zhou, New J. Phys. {\bf 12}, 025008 (2010); R. Or\'us and T.-C. Wei, Phys. Rev. B {\bf 82}, 155120 (2010).



\bibitem{Wei_Goldbart}T.-C. Wei and P. M. Goldbart, Phys. Rev. A \textbf{68}, 042307 (2003); T.-C. Wei, D. Das, S. Mukhopadyay, S. Vishveshwara, and P. M. Goldbart,
Phys. Rev. A {\bf 71}, 060305(R) (2005). 
\bibitem{brody}  D. C. Brody, L. P. Hughston, J. Geom. Phys. \textbf{38}, 19 (2001).
\bibitem{GGM} A. Sen(De) and U. Sen, Phys. Rev. A {\bf 81}, 012308 (2010); arXiv:1002.1253 [quant-ph]; 


\bibitem{CKW} V. Coffman, J. Kundu, and W. K. Wootters, Phys. Rev. A {\bf 61}, 052306 (2000); M. Koashi and A. Winter, Phys. Rev. A {\bf 69}, 022309 (2004); T. J. Osborne and F. Verstraete, Phys. Rev. Lett. {\bf 96}, 220503 (2006); G. Adesso, A. Serafini, and F. Illuminati, Phys. Rev. A {\bf 73}, 032345 (2006); T. Hiroshima, G. Adesso, and F. Illuminati, Phys. Rev. Lett. {\bf 98}, 050503 (2007); Y.-C. Ou and H. Fan, Phys. Rev. A {\bf 75}, 062308 (2007); M. Seevinck, ibid. {\bf 76}, 012106 (2007); S. Lee and J. Park, ibid. {\bf 79}, 054309 (2009); A. Kay, D. Kaszlikowski and R. Ramanathan, Phys. Rev. Lett. {\bf 103}, 050501 (2009);  M. Hayashi, and L. Chen, ibid. {\bf 84}, 012325 (2011);
F.F. Fanchini, M. C. de Oliveira, L.K. Castelano, and M. F. Cornelio, Phys. Rev. A {\bf 87}, 032317 (2013).

\bibitem{discmono} F. F. Fanchini, M. F. Cornelio, M. C. de Oliveira, and A.
O. Caldeira, Phys. Rev. A {\bf 84}, 012313 (2011);
R. Prabhu, A. K. Pati, A. Sen(De), and U. Sen, Phys. Rev. A {\bf 85}, 040102(R) (2012);
G. L. Giorgi, Phys. Rev. A {\bf 84}, 054301 (2011);
A. Streltsov, G. Adesso, M. Piani, and D Bruß, Phys. Rev. Lett. {\bf 109}, 050503 (2012);
K. Salini, R. Prabhu, A. Sen(De), and U. Sen, Ann. Phys. {\bf 348}, 297 (2014).
A. Kumar, R. Prabhu, A. Sen(De), U. Sen, Phys. Rev. A {\bf 91}, 012341 (2015).

\bibitem{Pan10photon}W. -B. Gao, C. -Y. Lu, X. -C. Yao, P. Xu, O. G\"uhne, A. Goebel, Y. -A. Chen, C. -Z. Peng, Z. -B. Chen, and J. -W. Pan, Nat. Phys. \textbf{6}, 331 (2010);
 T. E. Northup and R. Blatt, Nat. Photonics {\bf 8}, 356 (2014).


\bibitem{ION} D. Leibfried, R. Blatt, C. Monroe, and D. Wineland, Rev. Mod. Phys. {\bf 75}, 281 (2003); H. H\" affner, C. Roos, and R. Blatt, Phys. Rep. {\bf 469}, 155 (2008); T. Monz, P. Schindler, J. T. Barreiro, M. Chwalla, D. Nigg, W. A. Coish, M. Harlander, W. H\"ansel, M. Hennrich, and R. Blatt, Phys. Rev. Lett. {\bf 106}, 130506 (2011); J. T. Barreiro, J. -D. Bancal, P. Schindler, D. Nigg, M. Hennrich, T. Monz, N. Gisin and R. Blatt, Nat. Phys. \textbf{9}, 559 (2013).

\bibitem{mixed_ent} G. Gour, Phys. Rev. A {\bf 72}, 042318 (2005); C. Schmid, N. Kiesel, W. Wieczorek, H. Weinfurter, F. Mintert, and A. Buchleitner, Phys. Rev. Lett. {\bf 101}, 260505 (2008); M. Cianciaruso, T. R. Bromley, G. Adesso, arXiv:1507.01600 (2015).

\bibitem{GGM2}  R. Prabhu, S. Pradhan, A. Sen(De), and U. Sen, Phys. Rev. A {\bf 84}, 042334 (2011); H. S. Dhar, A. Sen(De), U. Sen, Phys. Rev. Lett. {\bf 111}, 070501 (2013); H. S. Dhar, A. Sen(De), and U. Sen, New J. Phys. {\bf 15}, 013043 (2013);
A. Biswas, R. Prabhu, A. Sen(De), and U. Sen, Phys. Rev. A {\bf 90}, 032301 (2014);
L. Jindal, A. D. Rane, H. S. Dhar, A. Sen(De), U. Sen, Phys. Rev. A {\bf 89}, 012316 (2014);  T. Das, R. Prabhu, A. Sen(De), and U. Sen, Phys. Rev. A {\bf 90}, 022319 (2014). 

\bibitem{Vollbrecht2}B. Terhal and K.G.H. Vollbrecht, Phys. Rev. Lett. {\bf 85}, 2625 (2000).


 

\bibitem{Vollbrecht}K. G. H. Vollbrecht and R. F. Werner, Phys. Rev. A {\bf 64}, 062307 (2001).


\bibitem{mixed_tangle}R. Lohmayer, A. Osterloh, J. Siewert and A. Uhlmann, Phys. Rev. Lett. {\bf 97}, 260502 (2006); C. Eltschka, A. Osterloh, J. Siewert and A. Uhlmann,
 New J. Phys. {\bf 10}, 043014  (2008).

\bibitem{symmetry_ent} 
T. Eggeling and R. F. Werner, e-print quant-ph/0003008;
W. D{\"u}r, J.I. Cirac, and R. Tarrach, Phys. Rev. Lett. {\bf 83}, 3562 (1999); R. Rains, quant-ph/0003008. 


\bibitem{GHZ} D. M. Greenberger, M. A. Horne, and A. Zeilinger, in Bell’s Theorem, Quantum Theory, and Conceptions of
the Universe, ed. M. Kafatos (Kluwer Academic, Dordrecht, The Netherlands, 1989).

\bibitem{Dicke} R. Dicke, Phys. Rev. {\bf 93}, 99 (1954).


 \bibitem{zhgstate} A. Zeilinger, M. A. Horne, and D. M. Greenberger, in Proceedings of Squeezed States and Quantum Uncertainty, 
                    edited by D. Han, Y. S. Kim, and W. W. Zachary, NASA Conf. Publ. \textbf{3135}, 73 (1992); W. D\"{u}r, G. Vidal, J. I. Cirac, Phys. Rev. A \textbf{62}, 062314 (2000); A. Sen(De), U. Sen, M. Wiesniak, D. Kaszlikowski, and M. \.Zukowski, Phys. Rev. A, 
                    \textbf{68}, 062306 (2003).
                    




                    
\bibitem{Bag_Mis_poster} T. Das, S. Singha Roy, S. Bagchi, A. Misra, A. Sen(De), and
U. Sen, poster presentation at the International Workshop on Quantum
Information, Harish-Chandra Research Institute, Allahabad, India, 20-26
February 2012, www.hri.res.in/~iwqi12/.

\bibitem{Otfried} L. E. Buchholz, T. Moroder and O. G\"uhne, arXiv:1412.7471 (2014).

 \end{thebibliography}
 \end{document}